\documentclass[
twocolumn
]{aastex631}
\usepackage{mathtools}
\usepackage{amsmath}

\shorttitle{Parametric decay of Alfv\'enic wave packets}
\shortauthors{Li, Fu \& Dorfman}

\graphicspath{{./}{figures/}}

\begin{document}

\title{Parametric decay of Alfv\'enic wave packets in nonperiodic low-beta plasmas}


\author[0000-0001-8638-8779]{Feiyu Li}
\affiliation{New Mexico Consortium, Los Alamos 87544, USA}

\author[0000-0002-4305-6624]{Xiangrong Fu}
\affiliation{New Mexico Consortium, Los Alamos 87544, USA}

\author[0000-0001-5955-9552]{Seth Dorfman}
\affiliation{Space Science Institute, Boulder, Colorado 80301, USA}
\affiliation{University of California Los Angeles, Los Angeles, California 90095, USA}

\begin{abstract}
The parametric decay of finite-size Alfv\'en waves in nonperiodic low-beta plasmas is investigated using one-dimensional hybrid simulations. Compared with the usual small periodic system, a wave packet in a large system under the absorption boundary condition shows different decay dynamics, including reduced energy transfer and localized density cavitation and ion heating. The resulting Alfv\'en wave dynamics are influenced by several factors of the instability including the growth rate, central wave frequency, and unstable bandwidth. A final steady state of the wave packet may be achieved when the instability does not have enough time to develop within the residual packet, and the packet size shows well-defined scaling dependencies on the growth rate, wave amplitude, and plasma beta. Under the proper conditions enhanced secondary decay can also be excited in the form of a narrow, amplified wavepacket. These results may help interpret laboratory and spacecraft observations of Alfv\'en waves, and refine our understanding of associated energy transport and ion heating. 
\end{abstract}


\section{Introduction} \label{sec:intro}

Alfv\'en waves (AWs) are of fundamental importance in magnetized plasmas. Due to their little geometrical attenuation, the waves may efficiently carry fluctuating magnetic energy over long distances, playing crucial roles in the energy transport found in space plasmas~\citep{de2007chromospheric}, astrophysical plasmas~\citep{jatenco1995alfven}, and fusion plasmas~\citep{chen2016physics}. However, at large amplitudes, the AWs may dissipate energy via nonlinear wave-wave and wave-particle interactions, which might be relevant to several important problems in space plasmas such as corona heating~\citep{heyvaerts1983coronal}, solar wind acceleration~\citep{suzuki2005making}, and minor ion heating~\citep{fu2020heating}. Parametric instabilities are a class of such nonlinear interactions that have gained considerable interest for the past half century~\citep{sagdeev1969nonlinear}. Dependent on the plasma beta (i.e., the ratio of thermal pressure to magnetic pressure), several instability types have been identified, including the decay, modulational, and beat instabilities~\citep{hollweg1994beat}. In low-beta plasmas, the parametric decay instability (PDI) predominates over the other two for its considerably higher growth rates. In the PDI, a forward pump AW decays to a backward AW and a forward sound wave. The generation of strong backward AWs in the PDI may trigger MHD (magnetohydrodynamic) turbulence via the interaction of counter-propagating AWs~\citep{goldreich1995toward}. The PDI may also strongly heat plasma through damping of the compressible sound wave~\citep[e.g.,][]{fu2018parametric,gonzalez2020role}.

The basic physics of PDI of AWs have been theoretically studied under various conditions. Sagdeev \& Galeev~(\citeyear{sagdeev1969nonlinear}) first analyzed the instability for linearly polarized AWs in the low-amplitude low-$\beta$ limit. Derby~(\citeyear{derby1978modulational}) and Goldstein~(\citeyear{goldstein1978instability}) extended the theory to finite wave amplitudes for circular polarization in the single-fluid MHD framework. Dispersive effects from ion cyclotron resonance were later included under the two-fluid framework, where the dispersion gives rise also to the modulational instability by enabling the interaction of forward propagating AW sidebands~\citep{sakai1983modulational,longtin1986modulation,wong1986parametric,hollweg1994beat}. The PDI process has also been studied via either MHD~\citep[e.g.,][]{ghosh1993parametric,del2001parametric,shi2017parametric}, hybrid~\citep[e.g.,][]{terasawa1986decay,vasquez1995simulation,araneda2007collisionless,matteini2010parametric,verscharen2012parametric,gonzalez2020role}, or particle-in-cell simulations~\citep[e.g.,][]{nariyuki2008parametric}, under different dimensions~\citep{ghosh1993parametric,del2001parametric}, driven by either monochromatic or broadband waves~\citep{malara1996parametric,shoda2018frequency}, coherent or incoherent waves~\citep{umeki1992decay}, with either parallel or oblique wave propagation~\citep{verscharen2012parametric,gao2013parametric}, single or multiple ion species~\citep{gao2013effects}, and in quiescent or turbulent background~\citep{shi2017parametric,fu2018parametric}. In the context of the Earth's foreshock and the solar wind, studies also show observational evidence of PDI~\citep[e.g.,][]{spangler1997observations,narita2007npgeo,dorfman2017foreshock,bowen2018density}, which has been simulated to illustrate, for example, the generation of density fluctuations~\citep{tanaka2007parametric}, the origin of low-frequency Alfv\'enic spectrum~\citep{reville2018parametric}, the effect of wind acceleration/expansion~\citep{shoda2018frequency}, and the evolution of switchbacks~\citep{tenerani2020magnetic}.

Despite these extensive studies, most of the theoretical treatment and simulations have essentially considered a periodic infinite system. This is achieved using the periodic boundary condition where the waves repeatedly interact with a localized plasma~\citep[e.g.,][]{del2001parametric,nariyuki2008parametric,gao2013parametric,shi2017parametric,tenerani2020magnetic,gonzalez2020role}. While this setup is useful for addressing the temporal dynamics of the instability, e.g., confirming the temporal growth rates~\citep{shi2017parametric}, actual AWs observed in space and laboratory are usually finite in time and space, and the associated interactions are nonperiodic~\citep{gigliotti2009generation}. These discrepancies may complicate the interpretation of laboratory or spacecraft observation of AW dynamics; for example, a modulational-like instability was recently observed in a laboratory plasma, while the PDI was mysteriously missing despite its significant growth rates in the low-beta environments~\citep{dorfman2016observation}. Pruneti \& Velli~(\citeyear{pruneti1997parametric}) first lifted the periodic-boundary constraint by considering wave injection into an open system using 1D/2D MHD simulations. Del Zanna et al.~(\citeyear{del2001parametric}) further extended it to 3D. However, these simulations involved only a few Alfv\'en wavelengths along the wave direction, thus were limited in capturing large-scale dynamics pertinent to the nonperiodic interactions. Notice that while the open boundary condition has been used in some solar-wind simulations to account for the global wind structure, the MHD framework is adopted to handle the vast space-time scales at the expense of ignoring kinetic effects~\citep[e.g.,][]{tanaka2007parametric,shoda2018frequency,reville2018parametric}.

The objective of this paper is to address how a finite-size AW packet decays in a large-scale nonperiodic system of uniform plasmas. This basic setup has fundamental interest and can also be realized in laboratory experiments with well-controlled plasma and wave conditions~\citep[e.g.,][]{gigliotti2009generation}. We consider a wave packet of tens of Alfv\'en wavelengths in size and examine its nonperiodic interaction with a low-beta plasma over a large time scale. This study is carried out using hybrid simulations with an absorption boundary condition for the wave fields. The large simulation domain considered here removes the constraints on the integer wave numbers as imposed under the periodic system. The hybrid simulation also allows us to keep the ion kinetic effects (e.g., wave damping, ion heating) that are absent from MHD simulations. 

The paper is organized as follows. In Section~\ref{sec:model}, we describe the simulation model, in particular the implementation of field masking for the absorption boundary condition. Novel PDI dynamics from the nonperiodic interactions are then detailed in Section~\ref{sec:results}, as well as its influences on the wave propagation. In Section~\ref{sec:discussion}, we discuss the results in relation to spacecraft and laboratory observations, and briefly mention future directions.

\section{Simulation model}\label{sec:model}

The simulations are performed with H3D, a three-dimensional hybrid code that self-consistently advances individual ion particles while treating electrons as a massless neutralizing fluid~\citep{karimabadi2006global}. Periodic boundary conditions were assumed by the code. In order to realize the absorption boundary condition, we apply a masking technique~\citep{umeda2001improved} to the field equations,
\begin{equation}
	\label{eq:ohm_law}
	    E=-u_i\times B-\frac{B\times(\nabla\times B)}{\mu_0 q_in_i}-\frac{\nabla P_e}{q_in_i}+\frac{\eta}{\mu_0}\nabla\times B, 
\end{equation}
\begin{equation}
    \label{eq:faraday_law}
	B^{n+1} = \overbrace{B^n-\Delta t(\nabla\times E)}^{\times F_m},
\end{equation}
where $E$ is the electric field, $B$ magnetic field, $\eta$ resistivity, $P_e$ electron pressure, and $u_i, n_i, q_i$  are the ion fluid velocity, density, and charge, respectively. The superscript $n$ represents the time step and $\Delta t$ is the step size. The mask consists of two layers on both sides of the $z$ boundaries ($z$ is along the mean magnetic field), and its profile $F_m$ is defined as 
\[
    F_m=\begin{dcases}
        1-\left(r\frac{z-\delta_m}{\delta_m}\right)^2, & z\leq \delta_m \\
        1, & \delta_m\leq z< L_z-\delta_m \\
        1-\left(r\frac{z-L_z+\delta_m}{\delta_m}\right)^2, & L_z-\delta_m \leq z \leq L_z\\
    \end{dcases}
\]
where $L_z$ is the domain size along $z$, $\delta_m$ the length of either mask layer, and $r$ a ratio factor controlling the slope of the mask. In H3D time subcycling is used for advancing the $B$ field with Equation~(\ref{eq:faraday_law}), thus the masking is applied only at the final subcycle. It is important that the background magnetic field $B_z$ is not masked. Equation~(\ref{eq:ohm_law}) is also left unmasked as the $E$ field will be automatically damped with $B$. It is found that the masking works significantly better with smaller $r$ (i.e., smaller gradients of $F_m$) but requires thicker mask layers to fully damp the waves, as was also observed for full Maxwell field solvers~\citep{umeda2001improved}. Using $r=0.1, \delta_m=5d_i$ we achieved 99\% absorption of the field energy upon $\sim 3d_i$ propagation into the mask layers, where $d_i=c/\omega_{pi}$ is the ion inertial length and $\omega_{pi}$ the ion plasma frequency. Notice that particles are still periodic, making the simulation less reliable at high wave amplitudes when strong plasma heating may induce fast particle circulation in the simulation domain. 

We carry out essentially 1D simulations by using one cell for the $x$ direction, four cells for $y$, and considerably more cells along the $z$ direction. The results have little dependence on $y$ and are therefore averaged over this coordinate. We take $z_{\rm max}=4400 d_i\equiv 110\lambda_0$ for simulations with the field masking, where $\lambda_0=2\pi v_A/\omega_0=40d_i$ is the Alfv\'en wavelength. In all simulations, the cell size is $\Delta x=\Delta y=\Delta z=1 d_i$ and time step $\Delta t=0.01\Omega_{ci}^{-1}=4\omega_{pi}^{-1}$ where $\Omega_{ci}$ is the ion cyclotron frequency satisfying $\omega_{pi}/\Omega_{ci}=400$. These parameters imply weakly dispersive AWs with $\omega_0/\Omega_{ci}=0.157$. A forward propagating (along $+z$) circularly-polarized AW packet of magnetic/velocity field
\begin{equation}
\begin{aligned}
    &\begin{aligned}
        \delta \vec{B}/B_0 &=  (\delta B_x/B_0)\chi(z)\sin(k_0z-\omega_0t)\hat{x} \\
        & -(\delta B_y/B_0)\chi(z)\cos(k_0z-\omega_0t)\hat{y},
    \end{aligned}\\
    &\delta \vec{u}=-v_A\delta\vec{B}/B_0,
\end{aligned}
\end{equation}
is initialized inside the domain next to the left mask layer (i.e., $z\geq \delta_m$), where $\delta B_x=\delta B_y=\delta B$ and $B_0=1$. The wave $E$ field is self-consistently determined from the field equations. The wave envelope $\chi(z)$ consists of sine-square-shaped ramps on sides and a plateau in between, i.e., 
\[
    \chi=\begin{dcases}
        \sin^2\left(\frac{\pi(z-\delta_m)}{2l_r}\right), & 0\leq z-\delta_m< l_r \\
        1, & 0\leq z-\delta_m-l_r<l_p \\
        \cos^2\left(\frac{\pi(z-z')}{2l_r}\right), & 0 \leq z-z'<l_r\\
    \end{dcases}
\]
where $z'=\delta_m+l_r+l_p$, and $l_r, l_p$ are the length of the ramp and plateau segment, respectively. The full packet size is $S_z=l_p+2l_r$. Notice that, despite the left-hand polarization, ion cyclotron resonance is not a concern here since $\omega_0/\Omega_{ci}\ll 1$. We have also run the simulation with right-hand polarization which gives very similar PDI dynamics. The plasma ions (protons) are sampled by 1000 macroparticles per cell. The initial density noise, $\delta\rho/\rho_0\sim 0.5\%$, is suitable for triggering the PDI over a reasonable time scale. We have confirmed very similar results using 64 particles per cell where the PDI develops slightly earlier due to larger initial noises. Electrons have the same initial temperature as ions and follow the adiabatic equation of state $T_e/n_e^{\gamma_e-1}=\rm const$ where $\gamma_e=5/3$. Therefore, the plasma beta is $\beta=\beta_i+\beta_e=2\beta_i$ where $\beta_i=\frac{n_iKT_i}{B_0^2/2\mu_0}=(\frac{v_i}{v_A})^2$ and $v_i=\sqrt{\frac{2KT_i}{m_i}}$ is the ion thermal speed. The comparable ion/electron temperatures may imply strong ion Landau damping. As we shall see, this kinetic effect mostly affects the nonlinear stage of PDI (i.e., when the sound wave grows to a high amplitude), leading to substantial ion heating. We have checked that the linear stage of PDI growth is well approximated by the fluid picture; the single-fluid MHD~\citep{derby1978modulational,goldstein1978instability} and two-fluid Hall MHD~\citep{hollweg1994beat} growth rates are also comparable due to the small $\omega_0/\Omega_{ci}$.

\section{Results}\label{sec:results}

\begin{figure*}[t]
\begin{center}
	\includegraphics[width=0.9\textwidth]{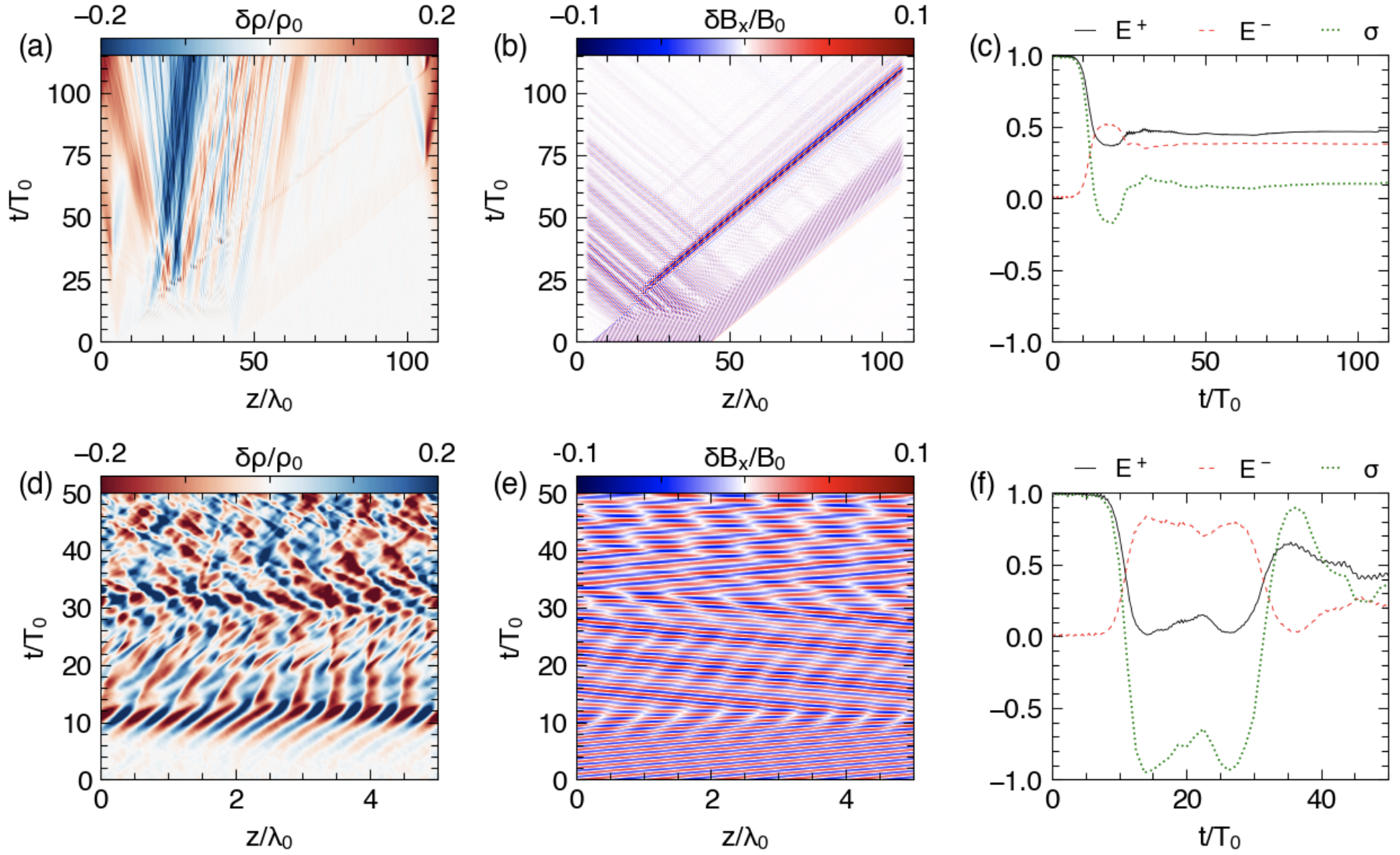}
	\caption{Space-time evolution of (a, d) density fluctuation $\delta\rho/\rho_0$ and (b, e) transverse $x$-component of the wave magnetic field $\delta B_x/B_0$. (c, f) Time evolution of forward wave energy $E^+$ (black solid), backward wave energy $E^-$ (red dashed), and cross helicity $\sigma$ (green dotted) where the wave energies are normalized to the initial value of $E^+$. The upper panels (a-c) are obtained for the absorption boundary condition and the bottom panels (d-f) for a periodic system. Detailed parameters are provided in the main text and notice the different axis ranges of the space-time plots.}
	\label{fig:open_vs_periodic}
\end{center}
\end{figure*}

Figure~\ref{fig:open_vs_periodic} first presents an overview of two contrast simulations, highlighting the difference between the more realistic absorption boundary condition (upper panels a-c) and the usual periodic system (lower panels d-f). In the former case,  a square wave packet of full size $S_z=40\lambda_0$ and small ramps $l_r=1\lambda_0$ is launched at $5\lambda_0\leq z\leq 45\lambda_0$. The latter periodic case has a smaller simulation domain, $z_{\rm max}=5\lambda_0$, which is fully occupied by a wave of constant amplitude initially. Both cases have the same wave amplitude $\delta B/B_0=0.1$ and plasma beta $\beta=0.02$. Significant differences are found in the space-time evolution of density fluctuation (a, d) and wave magnetic field (b, e). Under the absorption boundary condition the wave packet undergoes a strong decay at the early stage ($t<25T_0$ where $T_0=\lambda_0/v_A$) with the packet head being little distorted. Correspondingly, the density fluctuation is largely localized and constrained to $z<50\lambda_0$. In contrast, the periodic case shows mainly temporal dynamics where the density fluctuation grows or decays almost uniformly across the domain. 

To measure the strength of PDI, we separate the waves into forward- and backward-going components using the normalized Elsasser variables $z^\pm=\frac{1}{2v_A}(\delta u\mp\delta B/\sqrt{4\pi\rho})$, and define the associated wave energies and cross helicity as $E^\pm=\frac{1}{2}\sum |z^\pm|^2, \sigma=\frac{E^+-E^-}{E^++E^-}$, where the summation integrates over space and, in the absorption boundary case, includes also the absorbed waves in order to ensure energy conservation. By comparing between Figure~\ref{fig:open_vs_periodic}(c) and (f), it is seen that about 60\% of the pump energy ($E^+$) is decayed during the early stage in the absorption boundary case, while virtually the entire $E^+$ is depleted for the periodic system. This dramatic difference is also revealed by the evolution of the cross helicity which flips exactly in the latter case, indicating the dominance of backward waves as also seen in Figure~\ref{fig:open_vs_periodic}(e) during $15T_0<t<30T_0$. Despite this difference, about 80\% of the pump energy lost is transferred to the backward wave in both cases. To interpret that, we cast the frequency matching condition for the PDI in the quantum form $\hbar\omega_0=\hbar\omega+\hbar\omega_-$, which illustrates the energy apportioning between the sound wave ($\omega$) and backward lower-sideband AW ($\omega_-$). For $\beta=0.02$ in the weakly nonlinear regime, the mode of the maximum growth rate corresponds to $\omega/\omega_0=25\%$ and $\omega_-/\omega_0=75\%$~\citep{derby1978modulational}. The consistency in energy apportioning thus suggests that both cases contain the same PDI nature despite of very different dynamics. The two cases also share similar growth rates of the density wave (not shown), but their maximum density fluctuation amplitudes differ as the two start with different pump energies.

The absorption boundary case quickly evolves towards a steady state ($\sigma=0.1$ in this case) after the initial strong decay. However, the periodic system could undergo several flips of $\sigma$ until the wave energies in both directions exhaust. This latter phenomenon is apparently due to the repeated wave interaction with the local plasma forced by the periodic boundary condition. Thus, investigations using a periodic system will significantly overestimate the energy conversion from pump to daughter waves in the PDI, and are therefore of limited use in unraveling the PDI dynamics of finite-size AWs in more realistic nonperiodic systems.

\begin{figure*}[t]
\begin{center}
	\includegraphics[width=0.9\textwidth]{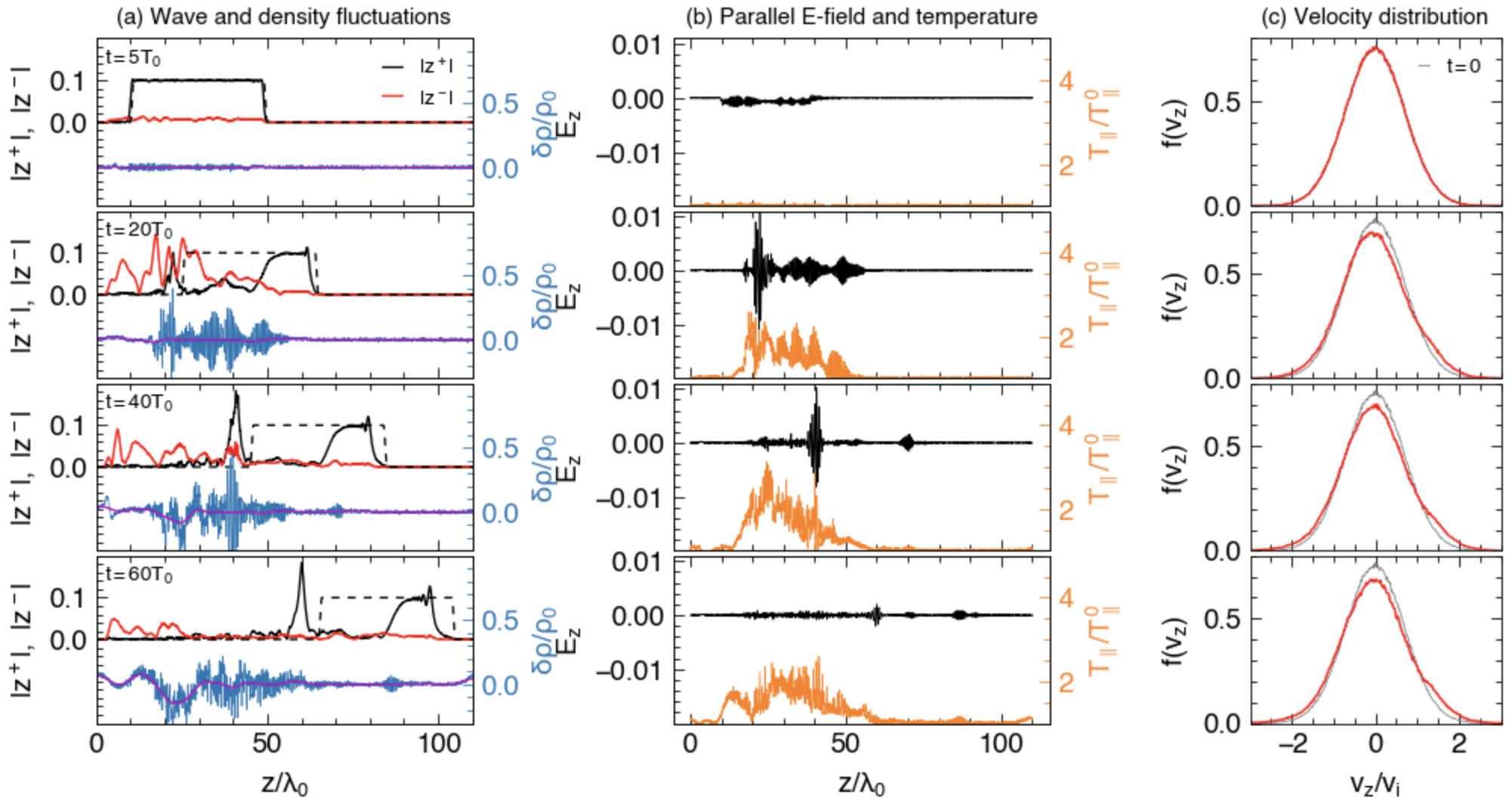}
	\caption{Snapshots of (a) the forward wave envelope $|z^+|$ (black), backward wave envelope $|z^-|$ (red), the initial wave envelope (translated to the current time step at the Alfv\'en speed, dashed line), density fluctuation $\delta\rho/\rho_0$ (blue), (b) parallel electric field $E_z$ (black) and parallel temperature $T_\parallel$ (orange), and (c) longitudinal ion velocity distribution (red) at $t=5, 20, 40, 60T_0$ (from top to bottom) for the absorption boundary case presented in Figure~\ref{fig:open_vs_periodic}. The magenta curves in (a) are smoothed density fluctuations, indicating any large-scale modulations. The grey curves in (c) represent the initial velocity distribution at t=0.}
	\label{fig:case_z110_tau40}
\end{center}
\end{figure*}

Let us now focus on the absorption boundary case and identify how the instability emerges, cascades, and saturates. Figure~\ref{fig:case_z110_tau40}(a) shows progressively the wave separation and associated density fluctuation. Initially, the wave is dominated by the square pump sitting close to the left boundary. The maximum growth rate for the present case is theoretically predicted to be $\gamma_{\rm max}\sim0.1\omega_0$. As a result, the PDI grows exponentially following $\exp(\gamma_{\rm max}t)$ in the linear stage. At $t=20T_0$, it is seen that the instability has already strongly developed: the back portion of the forward wave is much depleted, a significant backward wave appears, and a large-amplitude density fluctuation excited. Despite that, the pump front remains largely intact, but is gradually tailored to have a low-amplitude precursor and a smooth down-ramp profile towards the back. We term this entire front wave portion as the residual packet, because it stays rather stable against further propagation as shown at $t=40, 60T_0$. The formation of this steady state may be qualitatively understood as follows. In the frame comoving with the AW packet, its front portion is always fed with unperturbed plasmas. During the transit of the plasma through the wave, the density perturbation grows in a convective manner, becoming stronger at later times or towards the back of the wave. The density perturbation develops to sufficient levels once it reaches the position corresponding to the back end of the residual packet. This leads to the decay of the back portion of the pump wave, leaving only the residual packet intact.
Notice that the residual packet propagates at a speed smaller than the nominal Alfv\'en speed $v_A=B_0/\sqrt{\mu_0\rho_0}$ as seen by the delay relative to the dashed envelopes. The delay is found to be partly attributed to the finite-frequency effect, i.e., we obtain the group velocity $v_g=d\omega/dk_\parallel=(\omega/k_\parallel)/(1+k_\parallel^2d_i^2)\simeq 0.96v_A$ from the dispersion relation $\omega/k_\parallel=v_A\sqrt{1-(\omega/\Omega_{ci})^2}$ where $k_\parallel$ is the parallel wave vector along the background magnetic field. The wave bandwidth of the finite packet is among the other effects that would cause the additional delay.

Another remarkable feature is the emergence of a wave spike, trailing behind the pump (i.e., outside the assumed unperturbed envelope outlined by the dashed line). It has a large amplitude, twice that of the pump wave. This is the secondary decay, or inverse cascade as termed in some studies with the periodic system~\citep[e.g.,][]{del2001parametric}. It was  generated by a secondary PDI process where the backward AW serves as the pump. This secondary decay mostly occurs when the initial decay was strongly developed around $t=20T_0$. While secondary decays are also found in the periodic system, they never form such an isolated shape, involving both a large amplitude and a narrow duration/size. Thus, it may imply new potentials for secondary decays in real nonperiodic environments. The detailed mechanism of their formation is worthy of a separate study elsewhere.

Once the final steady state of the pump is reached, the PDI does not have enough time to develop as evidenced by the negligible density fluctuations ($\delta\rho/\rho_0<5\%$) excited within the modulated wave envelope. On the other hand, the large-amplitude density waves produced during the initial decay remain localized because of the small wave speed ($\sim 0.13v_A$) under the low-beta condition; see also the slope of the wave trace in Figure~\ref{fig:open_vs_periodic}(a). However, at later times, the density fluctuations gradually evolve into large-scale cavitation as indicated by the magenta curves in Figure~\ref{fig:case_z110_tau40}(a). This is caused by localized ion heating. To see that, Figure~\ref{fig:case_z110_tau40}(b) presents how the parallel electric field (black curves) develops with the compressible sound wave. Eventually the parallel wave field vanishes due to Landau damping of the sound wave. As a consequence, the ions are heated up to a parallel temperature $T_\parallel$ (orange) several times of the initial value $T_\parallel^0$, whereas the perpendicular temperature remains low throughout. The locally heated ions spread out longitudinally due to thermal expansion, causing the density cavitation. This long-term dynamics is also a reflection of the pressure balance, $P\propto nT$, where a higher-temperature region tends to have lower densities. The above heating effect is also evident from the velocity distribution as shown in Fig.~\ref{fig:case_z110_tau40}(c). While the distribution approximately maintains the Maxwellian-type profile, flattening of the distribution occurs between $v_z=[1, 2.2]v_i$. This is consistent with the sound wave speed $c_s=\sqrt{(\gamma_eKT_e+\gamma_iKT_i)/m_i}\simeq 1.3v_i$ considering $T_e=T_e$ and $\gamma_e=\gamma_i=5/3$.


The spatio-temporal (instead of just temporal) dynamics, along with the reduced energy transfer and localized density fluctuation/cavitation and ion heating, constitute our central new insights into the PDI process under the more realistic nonperiodic conditions. The spatial separation of Alfv\'enic fluctuations from density fluctuations and our new estimates of the efficiency of energy conversion and new insights on ion heating may have important implications for laboratory or spacecraft observations. For these applications, it may be imperative to understand better the Alfv\'enic propagation, especially how the final steady state depends on the plasma and wave conditions. 

\begin{figure}[t]
	\begin{center}
		\includegraphics[width=0.48\textwidth]{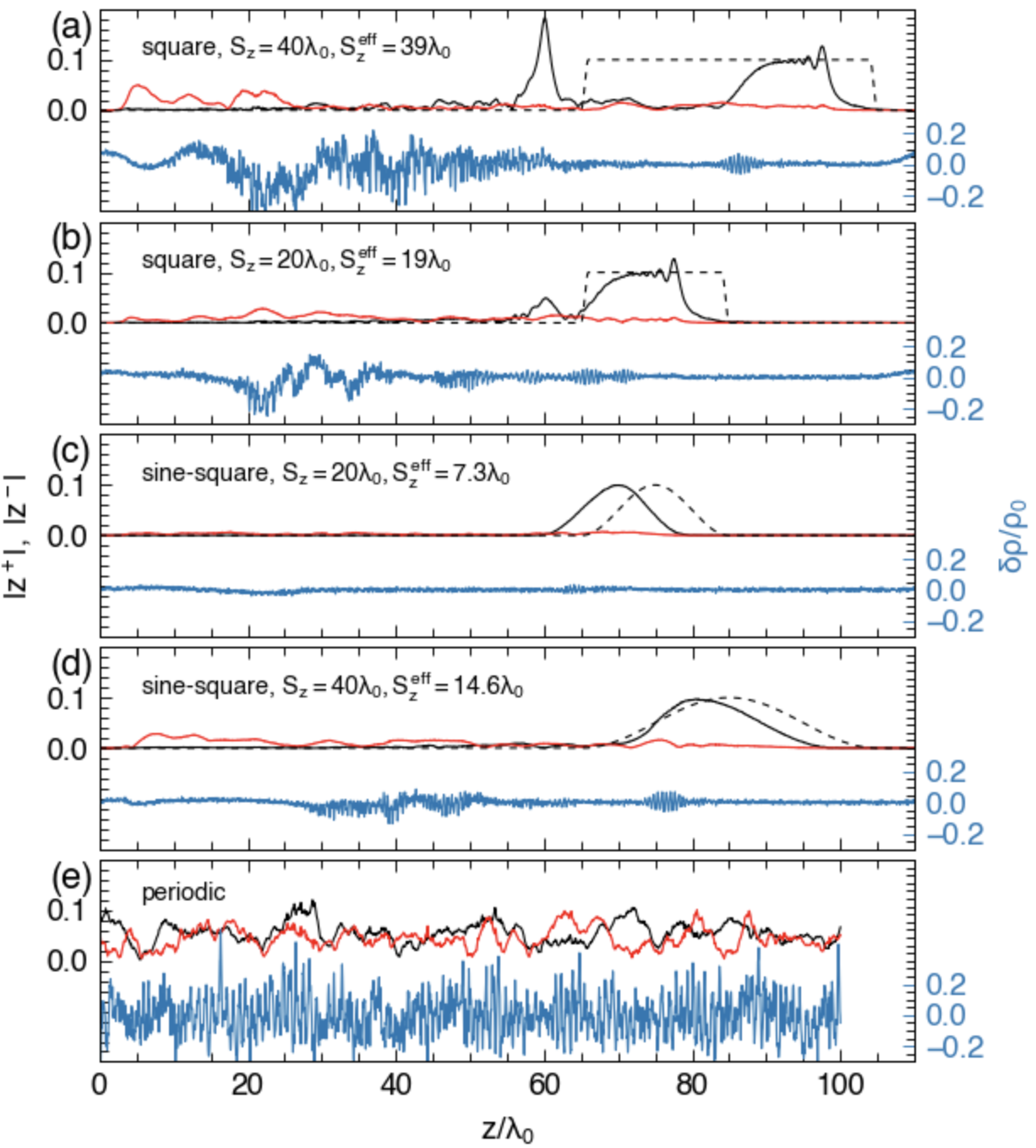}
		\caption{Snapshots of $|z^+|$ (black), $|z^-|$ (red), $\delta\rho/\rho_0$ (blue), and the initial wave envelope (translated to the current time step at the Alfv\'en speed, dashed line) for different cases (from top to bottom) at $t=60T_0$, where $S_z$ is the full packet size and $S_z^{\rm eff}$ the effective packet size measured at FWHM.}
		\label{fig:effect_pulse_duration}
	\end{center}
\end{figure}

\begin{figure*}[t]
\begin{center}
	\includegraphics[width=0.9\textwidth]{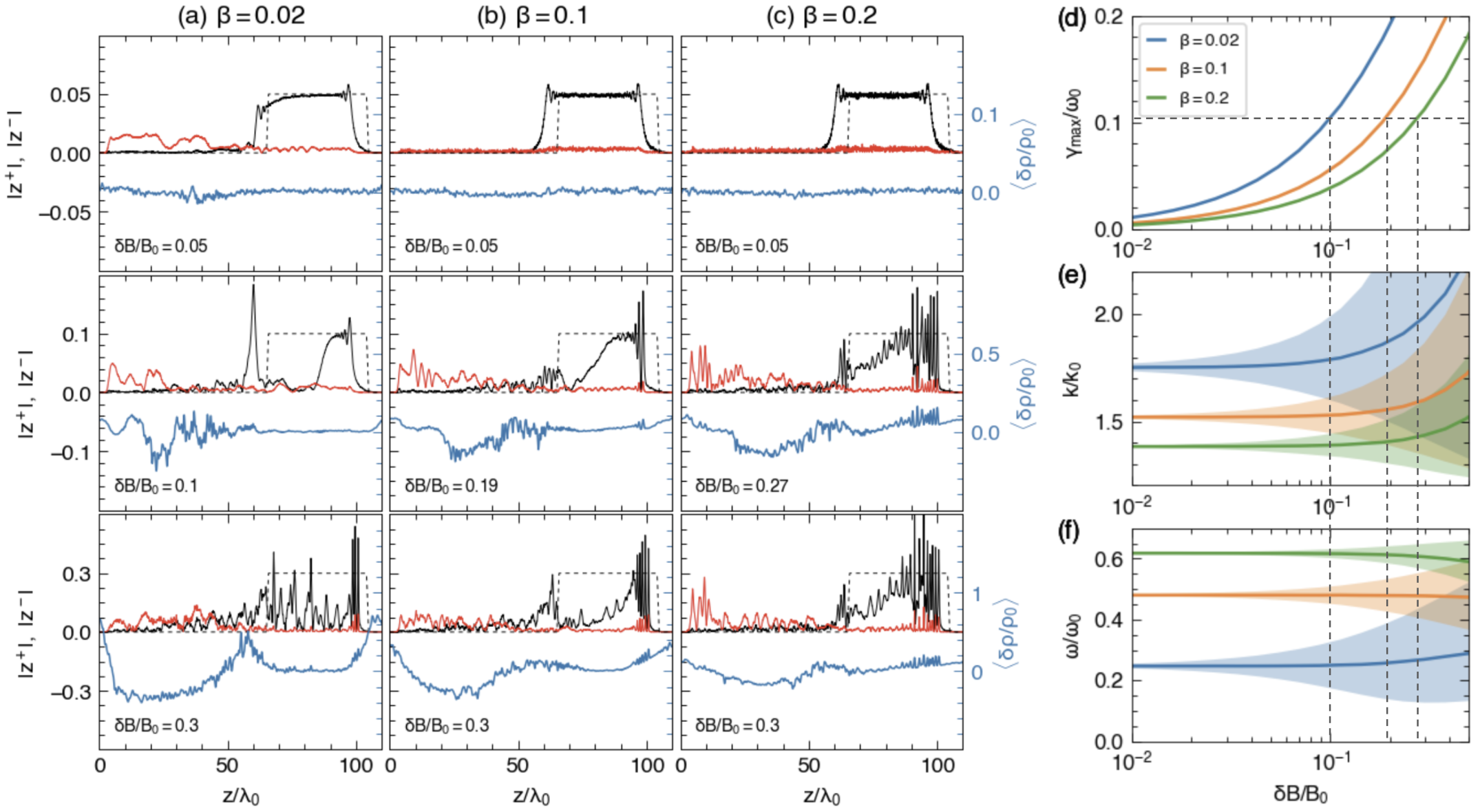}
	\caption{(a-c) Snapshots of $|z^+|$, $|z^-|$, and $\langle\delta\rho/\rho_0\rangle$ (smoothed over $1\lambda_0$ along $z$ axis) for cases of increasing $\delta B/B_0$ (from top to bottom) and of increasing $\beta$ (from left to right), shown at $t=60T_0$. The cases in the second row correspond to the same maximum growth rate (marked as dashed lines in d-f) despite their different $\delta B/B_0$ and $\beta$. (d) Maximum growth rates versus $\delta B/B_0$, solved from single-fluid MHD theory under different plasma beta $\beta$. (e) The wave number $k$ and (f) frequency $\omega$ of the compressible ion acoustic wave corresponding to the maximum growth rates versus $\delta B/B_0$. The shaded areas indicate the range within which the growth rates are non-negative.}
	\label{fig:parameter_scan}
\end{center}
\end{figure*}

As a first step, we investigate the effects of the wave envelope profile and packet size on the Alfv\'enic propagation. Figure~\ref{fig:effect_pulse_duration} displays the final steady states of several cases with different wave profiles or packet sizes (e.g., illustrated by the dashed lines). All of them are obtained under $\beta=0.02$ and $\delta B/B_0=0.1$. In the case of sine-square waves (i.e., no wave plateau region or $l_p=0$) the wave amplitude refers to the peak value. Case (a) is the baseline that we have discussed above, and it has an effective packet size (measured at full width at half maximum, FWHM) of $S_z^{\rm eff}=39\lambda_0$ initially. Compared with that, case (b) is only differed by a smaller packet size ($S_z^{\rm eff}=19\lambda_0$). We see virtually the same residual packet size for the two cases, $S_r^{\rm eff}\simeq 12\lambda_0$. This illustrates the irrelevance of initial packet size on the steady state so along as the initial packet is larger than its final residual packet, because the latter corresponds to only the front portion of the wave. However, larger initial packet sizes means more energies to be converted due to the PDI. This is evidenced by the stronger density fluctuations and more intense cascade spike found in case (a) as compared with (b). 
For an initial packet smaller than the final state size, no PDI should develop and the wave is stable throughout because of the short transit time. This is confirmed by case (c) where $S_z^{\rm eff}=7.3\lambda_0<12\lambda_0$. To clarify that the result is not due to the change of the initial wave envelope (i.e., from square to sine-square), we double the packet size [case(d)] and find that the sine-square wave becomes unstable again because its effective size $S_z^{\rm eff}=14.6\lambda_0>12\lambda_0$. These additional comparisons show the importance of packet size (or transit time), rather than the envelope, on the PDI excitation. The envelope may still play a role through the ponderomotive force, but its effect is generally small here due to the large packet size (or small envelope gradients) used. Finally, Figure~\ref{fig:effect_pulse_duration}(e) presents a periodic case having a wave size comparable to the simulation domain of the absorption boundary cases. Again, the instability develops uniformly across the domain, and no final steady state or enhanced wave cascades can be identified.

With the above insights, we next focus on the simple square wave profile and extend the investigation to a broad range of wave amplitude and plasma beta. The simulation results are summarized in Figure~\ref{fig:parameter_scan}(a-c) in terms of the final states and their corresponding density fluctuations. Furthermore, following the single-fluid theory~\citep{derby1978modulational,goldstein1978instability} which is justified here due to the small pump dispersion, the parameters $\delta B/B_0$ and $\beta$ adequately define the dispersion relation of the PDI, $\omega=\omega(k)$. The latter can be solved to find the growth rate, central frequency, and unstable bandwidth of the instability measured by the sound wave. These calculations, central to understanding the PDI and the resulting Alfv\'enic propagation, are presented in Figure~\ref{fig:parameter_scan}(d-f) in terms of the maximum growth rate $\gamma_{\rm max}$ and the bandwidth corresponding to positive $\gamma$. Of the most importance is the growth rate which defines the timescale relevant for the PDI development. In general, more significant decays happen with increasing growth rates, corresponding to larger $\delta B/B_0$ and lower $\beta$ following Figure~\ref{fig:parameter_scan}(d). This is verified by the overall pattern of the simulation results orderly arranged in Figure~\ref{fig:parameter_scan}(a-c). Closer to the top-right having smaller $\delta B/B_0$ and higher $\beta$, the wave is less prone to PDI. The trend reverses towards to bottom-left (having larger $\delta B/B_0$ and smaller $\beta$) where strong decays are found even at the early stage of the interactions. 

\begin{figure*}[t]
	\begin{center}
		\includegraphics[width=0.9\textwidth]{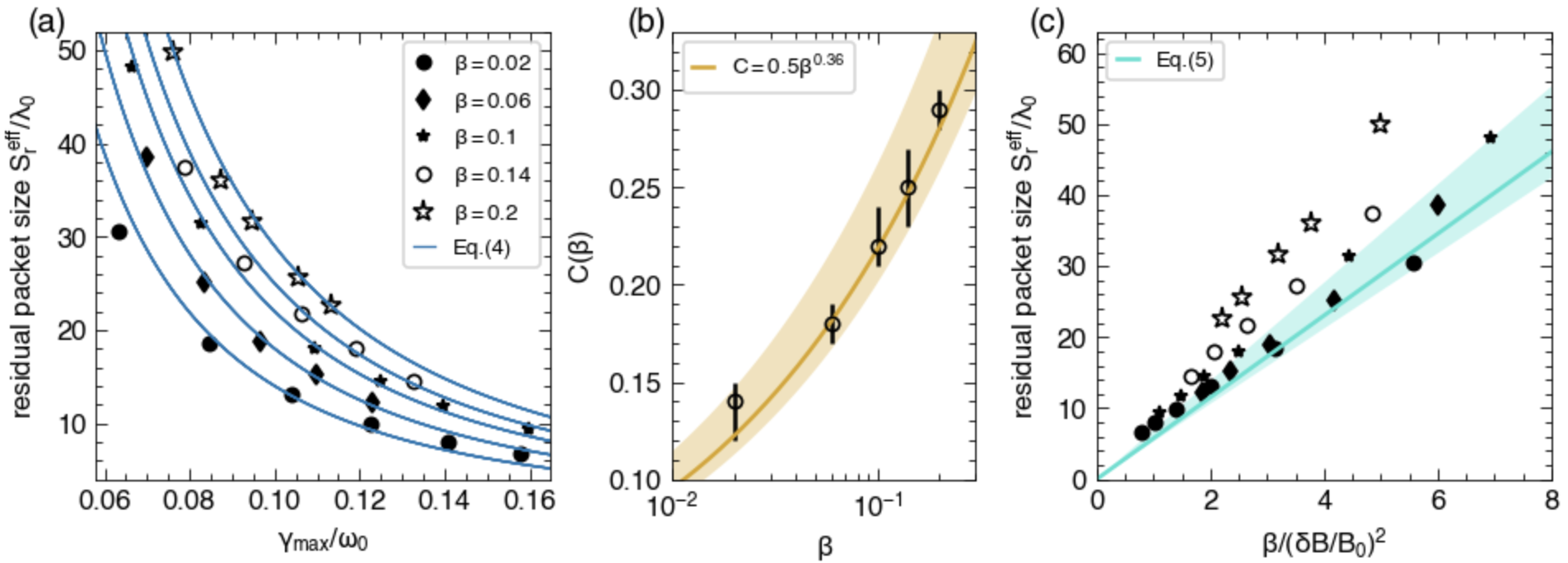}
		\caption{(a) The effective residual packet size, $S_r^{\rm eff}$, versus the maximum growth rate of the PDI under different $\beta$. All simulation results (point markers) are obtained with the square wave of an initial size $S_z=60\lambda_0$. The wave amplitudes $\delta B/B_0$ are chosen to have the maximum growth rate fall between $0.06\omega_0$ and $0.16\omega_0$. The solid curves represent fitting by Equation~(\ref{eq:residual_scaling}), where the coefficients $C(\beta)$ are presented in (b) as a function of $\beta$ (vertical lines indicate errorbars). The solid curve in (b) is a fitting of $C(\beta)$ using the $\beta^{0.36}$ dependence. (c) All simulation data presented in (a) are rearranged here using the combined axis $\beta/(\delta B/B_0)^2$. The solid line represents Equation~(\ref{eq:residual_scaling_beta_b}). The shaded areas in (b, c) come from the uncertainties in fitting $C(\beta)$.}
		\label{fig:residual_scaling}
	\end{center}
\end{figure*}

Despite its importance, the growth rate does not exclusively define the PDI-driven Alfv\'enic evolution. As seen in Figure~\ref{fig:parameter_scan}(e-f), the central $\omega$ and $k$ of the sound wave (solid curves) corresponding to the maximum growth rates also show strong dependence with $\beta$. For weakly nonlinear regimes (i.e., not too large $\delta B/B_0$), one has $\omega/\omega_0<1$ and $k/k_0\sim 2/(1+\sqrt{\beta})$, where $\omega/\omega_0$ increases and $k/k_0$ decreases with $\beta$, respectively. Therefore, following the energy apportioning relation $N\hbar\omega_-=N\hbar\omega_0-N\hbar\omega$, more energy goes into the backward AWs at smaller $\beta$ for a given pump depletion (i.e., fixed $N$ units of energy quanta). 
In addition, following $|k_-|=k-k_0$, one has $|k_-/k_0|\sim \frac{1-\sqrt{\beta}}{1+\sqrt{\beta}}<1$ which means $|k_-/k_0|$ keeps cascading into smaller numbers upon each secondary decay. The more limited $k$-space at higher $\beta$ (which corresponds to smaller initial $|k_-/k_0|$) may make secondary decays more difficult. This phenomenon was identified in periodic systems~\citep{del2001parametric} where the minimum wavenumber is limited by the box size.  It may also hold for the present nonperiodic condition when the packet size is large enough to allow for multiple decays. From the above analysis, both the frequency and wavenumber matching conditions suggest that the secondary wave decays favor smaller $\beta$ where more energy goes into the backward wave having larger $k_-/k_0$. This is supported by the simulation results where the secondary decays are stronger from column (c) to column (a) with decreasing $\beta$; for example, notice the prominent spike observed for $\beta=0.02, \delta B/B_0=0.1$. We have checked that the wave power in the low-frequency range (corresponding to the secondary decays) indeed increases with decreasing $\beta$ even in cases without a clearly visible cascade spike. 

A final contributing factor to the Alfv\'enic dynamics is the unstable mode bandwidth, i.e., the shaded areas in Figure~\ref{fig:parameter_scan}(e-f). The theory predicts rapidly increasing bandwidths with $\delta B/B_0$, becoming even larger than the central frequency/wavenumber as $\delta B/B_0\to 1$. The presence of multiple modes may cause strong competition among each other and make the wave envelopes more spiky. These features can be seen by comparing top to bottom panels in Figure~\ref{fig:parameter_scan}(a-c). 

The above analyses may point out the criteria for strong secondary decays with good wave coherency (e.g., narrow bandwidths and smooth envelopes). These phenomena would favor low-beta plasmas where the central frequency is small and the wavenumber is large, such that more energy goes into the backward waves with sufficiently large wavenumbers. Besides, the wave amplitude should not be too large, involving much mode competition and making the interactions less coherent. It should neither be too small requiring a large timescale for the PDI to develop.

Finally, we have assumed in the above analyses a certain pump depletion, which corresponds to the total energy lost to both the backward wave and sound wave. It can be defined as $\mathcal{D}=(S_z^{\rm eff}-S_r^{\rm eff})/S_z^{\rm eff}$. The quantity $1-D=S_r^{\rm eff}/S_z^{\rm eff}$ determines the portion of energy that can be stably transported over large distances and should be useful for estimating the energy transport efficiency and the limit of PDI-related plasma heating. As identified above, several factors may contribute to the Alfv\'enic dynamics, making analytical treatment of $D$ or $S_r^{\rm eff}$ a challenge. Here, we resort to simulations for its dependencies on the wave amplitude and plasma beta. We scan over a range of parameters where the maximum growth rate falls within $0.06<\gamma_{\rm max}/\omega_0<0.16$. Higher growth rates typically involve large $\delta B/B_0$ and broad mode competition, making the final state less distinguishable. Even lower growth rates may require a long time for PDI to develop sufficiently, a challenge to the simulations. Our results are summarized as point markers in Figure~\ref{fig:residual_scaling}(a). Despite the complexity of the interactions, a simple inverse-square dependence of the residual packet size versus the maximum growth rate is found as 
\begin{equation}
\label{eq:residual_scaling}
	\frac{S_r^{\rm eff}}{\lambda_0}=C(\beta)\left(\frac{\gamma_{\rm max}}{\omega_0}\right)^{-2},
\end{equation}
where the coefficient $C(\beta)$ has a power-law dependence on $\beta$, i.e., $C(\beta)\simeq 0.5\beta^{0.36}$, as displayed in Figure~\ref{fig:residual_scaling}(b). To more vividly illustrate the $\beta$-dependence, the second row of Figure~\ref{fig:parameter_scan}(a-c) shows three cases of different $\delta B/B_0$ and $\beta$ designed to have the same growth rate [marked as the dashed lines in Figure~\ref{fig:parameter_scan}(d-f)]. Despite the same growth rate, the final evolved states of the three cases are quite different, with significantly more pump depletion at smaller $\beta$.

In the low-amplitude low-beta limit, one may approximate the growth rate as $\gamma_{\rm max}/\omega_0\simeq \frac{1}{2.4\times \sqrt{2}}\frac{\delta B}{B_0}\frac{1}{\beta^{0.32}}$, an improved fitting from that suggested by Sagdeev \& Galeev~(\citeyear{sagdeev1969nonlinear}). By substituting it into Equation~(\ref{eq:residual_scaling}), we arrive at 
\begin{equation}
    \label{eq:residual_scaling_beta_b}
    \frac{S_r^{\rm eff}}{\lambda_0}=11.5C\frac{\beta^{0.64}}{(\delta B/B_0)^2}\simeq 5.8\frac{\beta}{(\delta B/B_0)^2},
\end{equation}
which explicitly associates the residual packet size with the plasma beta and wave amplitude. Notice a difference of $\sqrt{2}$ in wave amplitude when applying Sagdeev \& Galeev's scaling who studied linearly polarized AWs. We have verified Equation~(\ref{eq:residual_scaling_beta_b}) by arranging all simulation data in the combined axis $\beta/(\delta B/B_0)^2$ as displayed in Figure~\ref{fig:residual_scaling}(c). It is seen that they follow the linear scaling reasonably well, especially when the uncertainty in the coefficient of Equation~(\ref{eq:residual_scaling_beta_b}) is considered (the shaded area). The $\beta>0.1$ cases (i.e., the open points) are off, confirming the applicability of Equation~(\ref{eq:residual_scaling_beta_b}) to only the low-beta limit. In addition, high-amplitude cases typically involve broad mode bandwidths and spiky modulations of the wave envelope, making identification of the residual state more difficult.

\section{Discussion}\label{sec:discussion}

In this study, we have relaxed the usual periodic boundary condition and investigated the parametric decay dynamics of finite-size AWs in a large-scale low-beta plasma under the more realistic nonperiodic boundary condition. By simulating the dynamics over large space and time scales, our 1D hybrid simulations unravel novel dynamics that were not appreciated in the periodic system before, including strongly space-time dependent evolution, reduced pump energy transfer, and localized density fluctuation/cavitation and ion heating. These results may have important implications for spacecraft and laboratory observations. 

The robustness and importance of the PDI have stimulated the search of their existence in space over the past decades, for example, in the ultra-low-frequency wave field upstream of the Earth's bow shock~\citep{spangler1988observational,narita2007npgeo}. However, observational signature in the magnetic spectrum has been limited due to the dominance of the pump wave and the close separation of the pump and sidebands in frequency space~\citep{tsurutani1987steepened,agim1995magnetohydrodynamic}. The density spectral feature, on the other hand, may provide another independent signature of PDI because it is clearly separated from the magnetic signature in both wavenumber and frequency in low-beta plasmas~\citep{spangler1997observations}. 
As our simulations have shown, however, simultaneous observation of both magnetic and density signatures may not always be possible because the density fluctuation  can be separated in space and time from the Alfv\'enic fluctuation. This may explain the missing ``decay line'' signature reported in \cite{spangler1997observations} and help improve the search by establishing proper detection criteria. Furthermore, the  conjunction observation from multiple spacecraft in the heliosphere ~\citep{velli_a&a_2020} may identify missing signatures for PDI by correlating magnetic and density signatures at different locations along the same Parker spiral.

In addition, our simulations have shown that the front portion of the wave packet may evolve into a steady state under suitable wave and plasma parameters. This is because the instability has not enough time to grow in the wave packet. It also indicates efficient energy transport of a short wave train over long distances even in an environment where the PDI has a considerable growth rate. By combining simplified theories and numerical scans, we have identified well-behaved scaling dependencies of the residual packet size with both the growth rate, wave amplitude, and plasma beta. These scaling laws are valid for relatively low-beta low-amplitude regimes and should be useful for further analytical treatment on the energy transport. The AW propagation under the absorption boundary condition in higher-beta higher-amplitude regimes, which may apply to certain regions of the heliosphere~\citep{gary2001plasma}, is beyond the scope of the present work.

Finally, we have elucidated the major factors and corresponding parameter space that would influence the strength of decays and cascades. These insights should stimulate new interest in the basic PDI dynamics, especially in their highly nonlinear regime. For example, the PDI may modify the spectral evolution of inverse cascade~\citep{chandran2018parametric}. However, it should be pointed out that while our study has adopted a basic setup of fundamental importance, it is limited by the 1D geometry. High-dimensional effects such as finite wave $k_\perp$~\citep{dorfman2016observation}, oblique wave propagation~\citep{vinas1991parametric,verscharen2012parametric}, filamentation, magnetosonic decays, and perpendicular turbulence cascades~\citep{del2001parametric,gonzalez2020role} are among immediate next steps to check on the features identified in this work. We may also consider more realistic wave injection from the boundary for a finite period of time~\citep{del2001parametric}. The latter would have direct correspondence with laboratory experiments~\citep{gigliotti2009generation} and may help resolve some puzzles related to the experiments~\citep[e.g.,][]{dorfman2016observation}.

\vspace{0.2in}
This work is supported by NSF/DOE Partnership in Basic Plasma Science and Engineering under the grant DE-SC0021237. XF is also supported by the LANL/LDRD program and DOE/OFES. The authors acknowledge the Texas Advanced Computing Center (TACC) at The University of Texas at Austin and the National Energy Research Scientific Computing Center (NERSC) for providing HPC and visualization resources. NERSC is supported by the Office of Science of the U.S. Department of Energy under Contract No. DE-AC02-05CH11231.

\bibliography{pdi}{}
\bibliographystyle{aasjournal}

\end{document}